\documentclass[10pt]{iopart}
\usepackage{epsfig}
\usepackage{graphicx}
\newcommand{\be}{\begin{eqnarray}}
\newcommand{\ee}{\end{eqnarray}}

\newcommand{\ave}[1]{\left\langle #1 \right\rangle}

\begin{document}
\title{Fluctuations of particle ratios as a freeze-out probe}
\author{Giorgio Torrieri}
\address{Institut f\"ur Theoretische Physik,
  J.W. Goethe Universit\"at, Frankfurt A.M., Germany.}

\date{September, 2007}

\begin{abstract}
We explain how event-by-event fluctuations of particle ratios can constrain and falsify the statistical model of particle production in heavy ion collisions, using $K/\pi$ fluctuations as an example.
We define an observable capable of determining which statistical model, if any, governs freeze-out in ultrarelativistic heavy ion collisions.
We calculate this observable for $K/\pi$ fluctuations, and show that it should be the same for RHIC and LHC energies, as well as independent of centrality, if the Grand-Canonical statistical model is an appropriate description and chemical equilibrium applies.  We describe what happens in case of deviations from this scenario, such as light quark chemical non-equilibrium, strange quark over-saturation and local conservation (canonical ensemble) for strange quarks.   We also introduce a similar observable capable, together with the published $K^*/K$ measurement, of ascertaining if an interacting hadron gas phase governs the system between thermal and chemical freeze-out, and of ascertaining its duration and impact on hadronic chemistry
\end{abstract}
\maketitle

One of the main objectives of
heavy ion physics is to study the collective properties of strongly interacting matter.  It's equation
of state, degree of equilibration and phase structure, and the dependence of
these on energy and system size.
The natural approach to study particle production in heavy ion collisions is therefore statistical mechanics ~\cite{Fer50,Pom51,Lan53,Hag65}. In the last years, a consensus has developed that the statistical
hadronization model can indeed fit most or all particles yields for AGS,SPS and RHIC energies ~\cite{bdm,equil_energy,jansbook,becattini,nuxu}.

This consensus, is, however, somewhat superficial.  While it is true that one can get a reasonably nice-looking fit with a statistical model ansatz, it does not follow that the system is actually thermally and chemically equilibrated at freeze-out:  Considering the paucity of data points when particle abundances are modeled, such a fit is by itself not a guarantee of the physical significance of parameters such as temperature and chemical potential. When statistical significance of these fits is calculated, it is apparent that the statistical model is nowhere near ``proven'' according to the standards generally accepted in particle physics.
Particle fluctuations are a promising observable to falsify the statistical model and to constrain its parameters (choice of ensemble, strangeness/light quark chemical equilibrium) \cite{prcfluct}.
One can immediately see that fluctuations are a stringent statistical model test by considering the fluctuation of a ratio between two random variables.
\begin{equation}
\label{fluctratio}
\sigma_{N_1/N_2}^2
= \frac{\ave{(\Delta N_1)^2}}{\ave{N_1}^2}
+ \frac{\ave{(\Delta N_2)^2}}{\ave{N_2}^2}
- 2 \frac{\ave{\Delta N_1 \Delta N_2}}{\ave{N_1}\ave{ N_2}}.
\end{equation}
Since, for an equilibrated system, $\ave{(\Delta N_1)^2} \sim \ave{N} \sim \ave{V}$, where $\ave{V}$ is the system volume \cite{huang},
it is clear that $\sigma_{N_1/N_2}^2$ depends on the hadronization volume in a manner {\em opposite} to that of particle yields, inversely rather than directly linearly proportionally.    Volume fluctuations (which make a comparison of statistical model calculations to experimental data problematic), both resulting from dynamics and from fluctuations in collision geometry, should not alter this very basic result since volume cancels out event by event \cite{jeon}, {\em provided } hadronization volume is the same for all particles (a basic statistical model requirement).

  Thus, observables such as $\ave{N_1} \sigma_{N_1/N_2}^2$,
provided $\ave{N_{1,2}}$ and $\sigma_{N_1/N_2}^2$ are measured using the same kinematic cuts, 
 should be  strictly independent of multiplicity and centrality, as long as the statistical model holds and the physically appropriate ensemble is Grand Canonical.
If the temperature and chemical potentials between two energy regimes are approximately the same at freezeout (this should be the case for RHIC top energies and LHC, provided chemical equilibrium holds), this observable should also be identical.
This could be used as a stringent test of the statistical model.

Fluctuations are more sensitive to acceptance cuts than yields.
A partial ``fix'' for acceptance cuts that does not require detector-specific analysis is mixed event subtraction, based on the idea that fluctuation effects resulting from acceptance cuts are present both in real and mixed events (this is the case for fluctuations, but not for correlations).
Thus, an appropriate observable to model would be \cite{methods}
\begin{equation}
\sigma_{dyn}^2 = \sigma^2 - \sigma^2_{mix}
\end{equation}
where  $\sigma^2_{mix}$ is the mixed event width.    In the absence of any correlations, this reduces to the Poisson expectation. $\sigma_{dyn}^2$ (Known as \cite{methods} $\nu^{dyn}_{N_1/N_2}$ when measured using histograms rather than event-by-event) is currently subject of intense experimental investigation \cite{supriya,spsfluct}. We therefore propose to use \begin{equation}
\Psi^{N_1}_{N_1/N_2} = \ave{N_1} \nu_{dyn}^{N_1/N_2}
\end{equation}
to test the statistical model among different energy, system size and centralities.

``Primordial'' fluctuations of each observable, $\ave{(\Delta N_{1,2})^2}$, are calculable from Textbook methods \cite{huang}, but must also include corrections from resonance decays.  
Similarly, the correlation term in Eq. \ref{fluctratio} is also given, in the Grand Canonical ensemble, by resonance abundance 
\begin{equation}
\label{corrterm}
\ave{\Delta N_1 \Delta N_2} = \sum_j B_{j \rightarrow 1 2} \ave{N_j}
\end{equation}
SHAREv2.X \cite{share,sharev2} can calculate 
all 
ingredients of any $\Psi^{N_1}_{N_1/N_2}$, incorporating the effect of all resonance decays, as well as chemical (non)equilibrium.

It is important to underline that the value of $\Psi^{N_1}_{N_1/N_2}$ (calculated from statistical model parameters and, as we will see, sensitive to the degree of chemical equilibration of the system) should be {\em constant} across any system where the intensive parameters are the same, for example different centrality regimes or system sizes at the same energy.   For instance, since the chemical potential of Cu-Cu 200 GeV collisions should be comparable the chemical potential at Au-Au,  
\begin{equation}
\fl \left. \Psi^{\pi}_{K^-/\pi^-} \right|_{Cu-Cu} \simeq \left. \Psi^{\pi}_{K^-/\pi^-} \right|_{Au-Au}\phantom{AA}\left. \nu^{dyn}_{K^-/\pi^-} \right|_{Cu-Cu} \simeq \frac{\left.\ave{ N_\pi} \right|_{Cu-Cu}}{\left.\ave{ N_\pi} \right|_{Au-Au}}\left. \nu^{dyn}_{K^-/\pi^-} \right|_{Au-Au}
\end{equation}

A quantitative test for equilibrium statistical hadronizationcan also be made between RHIC and LHC energies.
Equilibrium thermal and chemical parameters are very similar at RHIC and the LHC(the baryo-chemical potential will be lower at the LHC, but, since both baryochemical potentials are small, the difference should not significantly affect $\pi$ and $K$ abundance).
Thus, $\Psi^{\pi^-}_{K^-/\pi^-}$ should be identical, to within experimental error, for both the LHC and RHIC, over all multiplicities were the statistical model is thought to apply.

According to \cite{janlhc}, chemical conditions at freeze-out (at SPS, RHIC and LHC) deviate from equilibrium, and reflect the higher entropy content and strangeness per entropy of the early deconfined phase through an over-saturated phase space occupancy for the light and strange quarks ($\gamma_{s}>\gamma_q>1$).
If this is true, than $\Psi^{N_1}_{N_1/N_2}$ should still be independent of centrality for a given energy range, but should go markedly up for the LHC from RHIC, because of the increase in $\gamma_q$ and $\gamma_s$.  

\begin{figure*}[h]
\begin{center}
\epsfig{width=12cm,figure=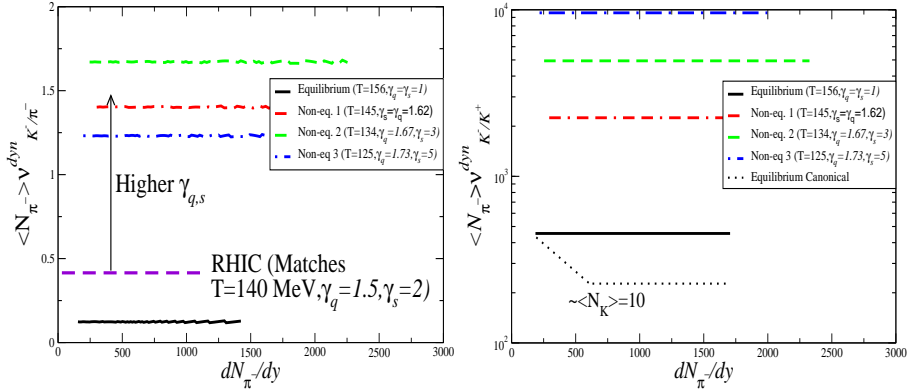}
\end{center}
\caption{\label{scalingplot} (color online) $\Psi^{\pi^-}_{K^-/\pi^-}$ (Left panel) and  $\Psi^{\pi^-}_{K^-/K^+}$ (right panel) calculated for various statistical hadronization parameters \cite{janlhc} at the LHC.  The left panel also shows the RHIC calculation \cite{sqm2006}   }
\end{figure*}

We have calculated $\Psi^{N_1}_{N_1/N_2}$ for RHIC and LHC energies, for the sets of parameters used in \cite{janlhc}.  The left and right panel in  Fig. 
\ref{scalingplot} shows what effect three different sets of $\gamma_{q,s}$ inferred in \cite{janlhc} would have on  $\Psi^{\pi^-}_{K^-/\pi^-}$ and $\Psi^{\pi^-}_{K^-/K^+}$.   In the left panel we have also included the value of  $\Psi^{\pi^-}_{K^-/\pi^-}$ for top energy RHIC.   As shown in \cite{sqm2006}, this value for top centrality matches expectations for non-equilibrium freeze-out (and is significantly above equilibrium freeze-out).   A centrality scan of  $\Psi^{\pi^-}_{K^-/\pi^-}$, necessary to confirm the consistency of this result has not, however, as yet been performed.

If non-statistical processes (mini-jets, string breaking etc.) dominate event-by-event physics, the flat $\Psi^{N_1}_{N_1/N_2}$ scaling on centrality/multiplicity should be broken, and $\Psi^{N_1}_{N_1/N_2}$ would exhibit a non-trivial dependence on $N_{part}$ or $dN/dy$.

This is also true if global correlations persist. such as is the case if the Canonical and micro-canonical ensembles \cite{gazdzicki} are physically more appropriate to describe the system.  Hence, if strangeness at RHIC/the LHC is created and maintained locally,  $\Psi^{N_1}_{N_1/N_2}$ should develop a ``wiggle'' at low centrality, and be considerably lower than Grand Canonical expectation. 
For  $\Psi^{\pi^-}_{K^+/K^-}$ it should be lower by a factor of two.


Fluctuations can also be useful in ascertaining the degree of reinteraction {\em after} an (assumed) statistical hadronization. A still unresolved ambiguity of statistical models
is the duration, and impact on hadronic observables, of the phase between hadronization (the moment at which particles become the effective degrees of freedom) and freeze-out (the moment at which particles stop interacting).

If, as commonly thought, chemical freezeout temperature is $ T_{chem} \sim 170$ MeV and thermal freezeout temperature is $T_{therm} \sim 100$ MeV, it follows that there is a significant interacting hadron gas phase that has the potential of altering all soft hadronic signatures.  The failure to solve problems associated with HBT \cite{hydroheinz}, however, suggests that something fundamental is still not understood about freeze-out, and more direct probes of freeze-out dynamics are needed.

The measurement of Resonance yields offers such a probe \cite{mereso1}, since short-lived hadronic resonances decay before the interacting hadron gas phase (if it exists) is over.
Thus, rescattering of decay products can deplete the amount of observable resonances, while regeneration could create additional resonances not present at hadronization.

The observation $\Lambda(1520)$ and $K^*(892)$ \cite{sevil}, at abundances below equilibrium statistical model expectations, could be interpreted as an indication of such reinteraction, with rescattering predictably dominating over regeneration.
This interpretation, however, is not unique: Chemical non-equilibrium fits recover the resonance abundance exactly, with no need for an interacting hadron gas phase
\cite{mereso1}.   
\begin{figure}[h]
\begin{center}
\epsfig{width=14cm,figure=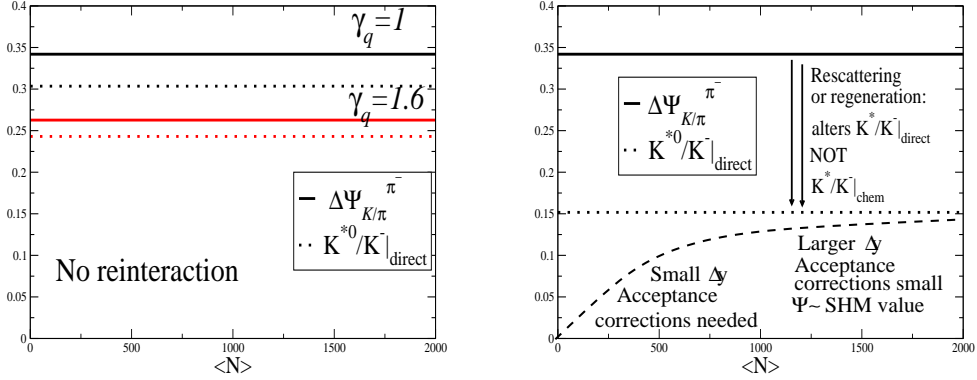}
\caption{\label{resoplot} (color online) $\Delta \Psi^{\pi^-}_{K/\pi}$ and $K^*/K^-$ calculated within the equilibrium and non-equilibrium statistical models}
\end{center}
\end{figure}

Thus detection of short lived resonances can not tell us freezeout dynamics unless a different signature, more sensitive to chemical freeze-out, is obtained.
As is apparent \cite{shuryak,jeon} from Eqs \ref{fluctratio} and \ref{corrterm}
the correlation term is precisely the required observable, since correlations between multiplicities are fixed at {\em chemical} freeze-out.
This correlation term can be measured by comparing observables such as $\Psi^{\pi^-}_{K^+/\pi^-}$ (correlated by $K^* (892)$) with  $\Psi^{\pi^-}_{K^-/\pi^-}$ (not correlated by resonances).   We therefore define
\begin{equation}
\fl \Delta \Psi^{\pi^-}_{K/\pi} = \frac{3}{4}  \ave{\pi^-} \left( \nu_{dyn}^{K^+/\pi^-} - \nu_{dyn}^{K^-/\pi^-}
\right) \simeq  \left. \frac{\ave{K^*(892) \rightarrow K^+ \pi^-}}{\ave{K^-}} \right|_{chemical\phantom{A}freeze-out}\phantom{AAAA}
\end{equation}
this result is somewhat spoiled by finite baryochemical potential, as well as higher lying resonances.   To ascertain the size of these corrections, we have used SHARE to calculate both the $K^*/K^-$ and $\Delta \Psi^{\pi^-}_{K/\pi}$.
As shown in Fig. \ref{resoplot} (left panel), these corrections make up a 10 $\%$ effect, less than the expected experimental error and not enough to alter the difference between a single freeze-out and two simultaneous ones.

A long reinteracting hadron gas phase would in general bring the observed (final) 
abundance of $K^*/K$ away from the chemical freezeout value (either up, by regeneration, or down, by rescattering).   Thus, $\Delta \Psi^{\pi^-}_{K/\pi}$
would become different from $K^*/K^-$ (Fig. \ref{resoplot} right panel).
In the weak interaction limit, regeneration would presumably be rarer than rescattering so the observable  $K^*/K^-$ abundance would be suppressed by a factor that combines the interaction width $\Delta \Gamma=\Gamma_{rescattering}-\Gamma_{regeneration}$ with the lifetime of the interacting phase $\tau$
\begin{equation}
\left. \frac{K^*}{K^-} \right|_{observed} \sim \Delta \Psi^{\pi^-}_{K/\pi} \exp \left[ -\Delta \Gamma \tau \right]
\end{equation}
In the strong reinteraction limit, rescattering and regeneration would re-equilibrate to a lower freeze-out $T_{therm}$, so the observed $K^*/K^-$ would be sensitive to the temperature difference as well as the mass difference ($\Delta m$) between $K^*$ and $K$
\begin{equation}
\left. \frac{K^*}{K^-} \right|_{observed} \sim \Delta \Psi^{\pi^-}_{K/\pi} \exp \left[ \frac{\Delta m}{T_{therm}} -  \frac{\Delta m}{T_{chem}} \right]
\end{equation}
One untested effect that could spoil this result is strong rescattering capable of bringing the $K$ and $\pi$ out of the detector's acceptance region in phase space \cite{prcfluct,jeon}.
This effect can not be taken into account by mixed event techniques described in the previous section, and calculating it in a model-independent way is problematic.

Inferring the presence of such a correction is however relatively straight-forward (Fig. \ref{resoplot},right panel): The probability of such rescattering strongly depends on the width of the acceptance region.
Thus, if $\Delta \Psi^{\pi^-}_{K/\pi}$
 as a function of the rapidity window should go from zero (at small rapidity no multiplicity correlation survives) and saturate at a constant value (where the full resonance derived correlation is recovered).  This constant value, as long the rapidity window is much smaller than the total extent of rapidity of the system, is the quantity that can be related to $\left. \frac{K^*}{K^-} \right|_{chem}$.   

The dependence on centrality of $\Delta \Psi^{\pi^-}_{K/\pi}$, on the other hand, has to remain flat, since in the Grand Canonical limit the ratio of two particles should be independent of centrality, and the total system size should not alter the probability of a local process (such as scattering in/out of the acceptance region) to occur.  

 If $\Delta \Psi^{\pi^-}_{K/\pi}$ obeys the consistency checks described here
it should be taken as an indication that the $\Delta \Psi^{\pi^-}_{K/\pi}$
 measurement reflects the value of  $\frac{K^*}{K^-}$ at chemical freezeout.  

If $\Delta \Psi^{\pi^-}_{K/\pi}$ depends on rapidity up to the acceptance region of the detector, a more thorough effort to account for acceptance corrections to the correlation is needed.   This can be done by using the same techniques utilised for direct resonance reconstruction \cite{sevil}.
Such an endeavour is detector specific, and goes beyond this write-up.

In conclusion, we have shown that observables incorporating both yields and fluctuations give a stringent test of statistical models.
We have also argued that such observables can be used to infer the duration of the interacting hadron gas phase, and its effect on hadronic observables.
We expect that forthcoming experimental data, together with the methods elucidated here, will allow us to clarify some of the outstanding puzzles apparent in the study of heavy ion collisions.

GT thanks the Alexander Von Humboldt
foundation, the Frankfurt Institute for Theoretical Physics and FIAS for continued support.
We would also like to thank theorganizing committee for allowing me to attend SQM2007 at such short notice.

\end{document}